\documentclass[twocolumn]{aastex631}

\usepackage{graphicx}		
\usepackage{latexsym, float}		
\usepackage{bm}			
\usepackage[english]{babel}
\usepackage{lipsum}
\usepackage{natbib}
\usepackage{appendix}

\usepackage{siunitx}
\usepackage{booktabs}
\usepackage{enumitem}
\usepackage{color}		
\usepackage{comment}

\newcommand{\lsim}{\mbox{\rlap{\hbox{\lower2pt\hbox{\ensuremath{\sim}}}}\raise2pt\hbox{\ensuremath{<}}}}%
\newcommand{\gsim}{\mbox{\rlap{\hbox{\lower2pt\hbox{\ensuremath{\sim}}}}\raise2pt\hbox{\ensuremath{>}}}}%
\renewcommand{\gtrsim}	{\ensuremath{\gsim}}
\renewcommand{\lesssim}	{\ensuremath{\lsim}}
\newcommand\eg           {{\it e.g.}, }

{\relax}%
\newcommand\n            {\noindent}

\newcommand\degree       {{\ifmmode^\circ\else$^\circ$\fi}} 
\newcommand\arcm         {{\ifmmode {'\ }\else$'     $\fi}} 
\newcommand\arcs         {{\ifmmode{''\ }\else$''    $\fi}} 
\newcommand{\DELETED}[1]{\relax}%
{\relax}%

\usepackage{lineno}
\usepackage{xurl}
\begin{document}
\title{Globular Clusters in the Galaxy Cluster MACS0416 at $z=$ 0.397}

\author[0000-0001-6265-0541]{Jessica M. Berkheimer}
\affiliation{School of Earth \& Space Exploration, Arizona State University, Tempe, AZ\,85287-6004, USA}
\affiliation{Beus Center for Cosmic Foundations, Arizona State University, Tempe, AZ\,85287-6004, USA}

\author[0000-0001-8156-6281]{Rogier A.~Windhorst} 
\affiliation{School of Earth \& Space Exploration, Arizona State University, Tempe, AZ\,85287-6004, USA}

\author[0000-0001-8762-5772]{William E. Harris}
\affiliation{Department of Physics \& Astronomy, McMaster University, 1280 Main Street West, Hamilton, L8S 4M1, Canada}

\author[0000-0002-6610-2048]{Anton M.~Koekemoer} 
\affiliation{Space Telescope Science Institute, 3700 San Martin Drive, Baltimore, MD\,21218, USA}

\author[0000-0001-6650-2853]{Timothy Carleton}
\affiliation{School of Earth \& Space Exploration, Arizona State University, Tempe, AZ\,85287-6004, USA}

\author[0000-0003-3329-1337]{Seth H.~Cohen} 
\affiliation{School of Earth \& Space Exploration, Arizona State University, Tempe, AZ\,85287-6004, USA}

\author[0000-0003-1268-5230]{Rolf A.~Jansen} 
\affiliation{School of Earth \& Space Exploration, Arizona State University, Tempe, AZ\,85287-6004, USA}

\author[0000-0001-7410-7669]{Dan Coe} 
\affiliation{AURA for the European Space Agency (ESA), Space Telescope Science Institute, 3700 San Martin Drive, Baltimore, MD\,21218, USA}

\author[0000-0001-9065-3926]{Jose Diego} 
\affiliation{Instituto de Física de Cantabria (CSIC-UC). Avda. Los Castros s/n. 39005 Santander, Spain}

\author[0000-0003-1949-7638]{Christopher J. Conselice}
\affiliation{Jodrell Bank Centre for Astrophysics, Alan Turing Building, University of Manchester, Oxford Road, Manchester M13 9PL, UK}

\author[0000-0001-9491-7327]{Simon P. Driver} 
\affiliation{International Centre for Radio Astronomy Research (ICRAR) and the International Space Centre (ISC), The University of Western
Australia, M468, 35 Stirling Highway, Crawley, WA 6009, Australia}

\author[0000-0003-1625-8009]{Brenda L.~Frye} 
\affiliation{Department of Astronomy\,/\,Steward Observatory, University of Arizona, 933 N.\ Cherry Ave., Tucson, AZ\,85721, USA}

\author[0000-0001-9440-8872]{Norman A.~Grogin} 
\affiliation{Space Telescope Science Institute, 3700 San Martin Drive, Baltimore, MD\,21218, USA}

\author[0000-0001-5290-6275]{Kate Hartman}
\affiliation{Department of Physics \& Astronomy, McMaster University, 1280 Main Street West, Hamilton, L8S 4M1, Canada}

\author[0009-0008-0376-3771]{Tyler R. Hinrichs}
\affiliation{School of Earth and Space Exploration, Arizona State University, Tempe, AZ 85287-6004, USA
}

\author[0000-0002-4884-6756]{Benne W. Holwerda} 
\affiliation{Department of Physics and Astronomy, University of Louisville, Louisville, KY 40292, USA} 

\author[0000-0001-9394-6732]{Patrick S. Kamieneski}
\affiliation{School of Earth and Space Exploration, Arizona State University, Tempe, AZ 85287-6004, USA
}

\author[0009-0001-7599-1967]{Kaitlyn E. Keatley}
\affiliation{Department of Physics \& Astronomy, McMaster University, 1280 Main Street West, Hamilton, L8S 4M1, Canada}

\author[0000-0002-6131-9539]{William C. Keel} 
\affiliation{Department of Physics and Astronomy, University of Alabama, Box 870324, Tuscaloosa, AL\,35404, USA}

\author[0000-0003-1581-7825]{Ray A. Lucas} 
\affiliation{Space Telescope Science Institute, 3700 San Martin Drive, Baltimore, MD\,21218, USA}

\author[0000-0001-6434-7845]{Madeline A.~Marshall} 
\affiliation{National Research Council of Canada, Herzberg Astronomy \& Astrophysics Research Centre, 5071 West Saanich Road, Victoria, BC V9E\,2E7, Canada}

\author[0000-0001-6342-9662]{Mario Nonino} 
\affiliation{INAF-Trieste Astronomical Observatory, Via Bazzoni 2, 34124, Trieste, Italy}

\author[0000-0003-3382-5941]{Nor Pirzkal} 
\affiliation{Space Telescope Science Institute, 3700 San Martin Drive, Baltimore, MD\,21218, USA}

\author[0000-0003-4223-7324]{Massimo Ricotti}
\affiliation{Department of Astronomy, University of Maryland, College Park, 20742, USA}

\author[0000-0002-5404-1372]{Clayton D. Robertson} 
\affiliation{Department of Physics, University of Louisville, Natural Science Building 102, 40292 KY Louisville, USA}

\author[0000-0003-0429-3579]{Aaron Robotham}
\affiliation{
International Centre for Radio Astronomy Research (ICRAR) and the International Space Centre (ISC), The University of Western
Australia, M468, 35 Stirling Highway, Crawley, WA 6009, Australia}

\author[0000-0003-0894-1588]{Russell E.~Ryan, Jr.} 
\affiliation{Space Telescope Science Institute, 3700 San Martin Drive, Baltimore, MD\,21218, USA}

\author[0000-0002-7265-7920]{Jake Summers}
\affiliation{School of Earth \& Space Exploration, Arizona State University, Tempe, AZ\,85287-6004, USA}

\author[0000-0001-9262-9997]{Christopher N.A.~Willmer} 
\affiliation{Department of Astronomy\,/\,Steward Observatory, University of Arizona, 933 N.\ Cherry Ave., Tucson, AZ\,85721, USA}

\author[0000-0001-7592-7714]{Haojing Yan}
\affiliation{Department of Physics and Astronomy, University of Missouri, Columbia, MO\, 65211, USA}

\correspondingauthor{Jessica Berkheimer}
\email{jberkhei@asu.edu}

\shortauthors{Berkheimer et. al}
\shorttitle{MACS0416}

\begin{abstract}
We present a photometric analysis of globular clusters (GCs) in the massive galaxy cluster MACS J0416.1–2403 ($z = 0.397$) using deep JWST/NIRCam imaging from the PEARLS program. PSF photometry was performed in the short wavelength filters, $F090W, F115W, F150W,$ and $F200W$, yielding a catalog of $ \sim3\times10^3$ unresolved, point-like sources consistent with a GC population. Artificial-star tests indicate 80\% completeness at $F200W \simeq 30.36$~mag. 
The color--magnitude diagrams show a narrow GC sequence well reproduced by PARSEC single-stellar-population models spanning ages of 5--9~Gyr and metallicities from [M/H]~$\approx -2.0$ to $+0.2$, consistent with evolved GC systems at this redshift. 
The globular cluster luminosity function (GCLF) follows a log-normal form truncated by incompleteness at the faint end. 
The brightest sources extend slightly beyond the locus of classical GCs, suggesting a small number of UCD-like systems or stripped nuclei, while the bulk of the population exhibits the luminosities and colors expected for mature globular clusters at $z \simeq 0.4$. 

\end{abstract} \hspace{12pt}


\section{Introduction}

GCs are among the most ancient stellar systems in the universe. These tightly bound collections of up to a million stars are thought to trace the early conditions of galaxy formation, making them valuable fossil records of the assembly history of galaxies \citep[\eg][]{cote1998formation, carretta2000distances, krauss2003age, marin2009acs, kruijssen2015globular,kruijssen2025,beasley2020globular}. GCs are found in galaxies across the full mass spectrum, from dwarfs to the most massive ellipticals in galaxy clusters \citep[\eg][]{brodie2006extragalactic, harris2013catalog, harris2023photometric,  harris2024jwst}. Due to their intrinsic brightness and compactness, GCs can be observed with JWST as point sources at cosmological distances, even against the brighter background from cluster galaxies or the intracluster light (ICL) \citep[e.g.,][]{vanzella2023jwst, senchyna2024gn}.

MACS~J0416.1--2403 (hereafter MACS0416) is a rich, lensing galaxy cluster at $z = 0.397$. Its strong gravitational lensing and complex mass distribution have made it a key target in several major surveys, including the Hubble Frontier Fields (HFF), CLASH, RELICS, and, most recently, the PEARLS program with \textit{JWST} \citep[\eg][]{diego2015free, lotz2017frontier, bergamini2023state, windhorst2023jwst}. These observations have revealed hundreds of multiply-imaged lensed sources and provided detailed constraints on the cluster’s dark matter distribution and its sub-structure \citep[\eg][]{zitrin2012clash, grillo2015clash, diego2023jwst, perera2024buffalo, rihtarvsivc2025canucs}, an important factor in interpreting its globular cluster population.

MACS0416 has a projected mass of $M \sim 3-4 \times 10^{14}\,M_\odot$ within $R \sim 400$~kpc, based on lensing constraints from the inner cluster region \citep[\eg][]{balestra2016clash, diego2023jwst}. However, full lensing and X-ray analyses place the total virial mass of the cluster at $M_{\rm vir} \sim 1-1.4 \times 10^{15}\,M_\odot$ when measured within $R_{200} \sim 2-2.5$~Mpc \citep[\eg][]{umetsu2014clash, zitrin2013clash}. With this total mass, MACS0416 ranks among the most massive galaxy clusters known at $z \sim 0.4$, making it a valuable environment for studying the formation and evolution of GC systems in high-mass halos.

\textit{JWST}'s deep multiband imaging enables unprecedented resolution in detecting and analyzing individual GCs in MACS0416.

In this study, we present a photometric analysis of GCs in MACS0416 using high-resolution NIRCam short-wavelength (SW) imaging from the PEARLS program. Section~\ref{sec:data} describes the observations, PSF characterization, and photometric procedures. Section~\ref{sec:completeness} quantifies completeness using artificial-star tests. Section~\ref{sec:colors} presents the GC color distributions and statistical analysis of subpopulations. Section~\ref{sec:gclf} discusses the construction and modeling of the GCLF, while Section~\ref{sec:kcorr} applies $K$-corrections. Section~\ref{sec:ucd_comparison} and Section~\ref{sec:stellar_models} extend the analysis to the physical interpretation of the cluster population, examining GC mass scales, the possible connection to ultra-compact dwarfs (UCDs) and stripped nuclei, and consistency with single-stellar-population (SSP) models. We adopt AB magnitudes and assume a flat $\Lambda$CDM cosmology with $H_0 = 67.8\,\mathrm{km\,s^{-1}\,Mpc^{-1}}$, $\Omega_\Lambda = 0.69$, and $\Omega_m = 0.31$ \citep{ade2016planck}. At the redshift of MACS0416 ($z = 0.397$), this corresponds to a luminosity distance of $d_L = 2216.5$~Mpc, a distance modulus of $(m\!-\!M)_0 = 41.73$, and a lookback time of 4.38~Gyr. The angular diameter distance is $d_A = 1135.1$~Mpc, yielding a physical scale of 5.5~kpc per arcsecond. The small but non-negligible foreground Galactic extinction values (Table~\ref{tab:kcorr}) have been corrected for in all photometry.

\begin{figure*}[ht]
\centering
\includegraphics[width=0.48\textwidth]{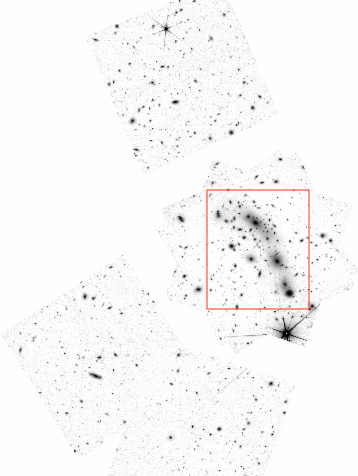}
\hfill
\includegraphics[width=0.5\textwidth]{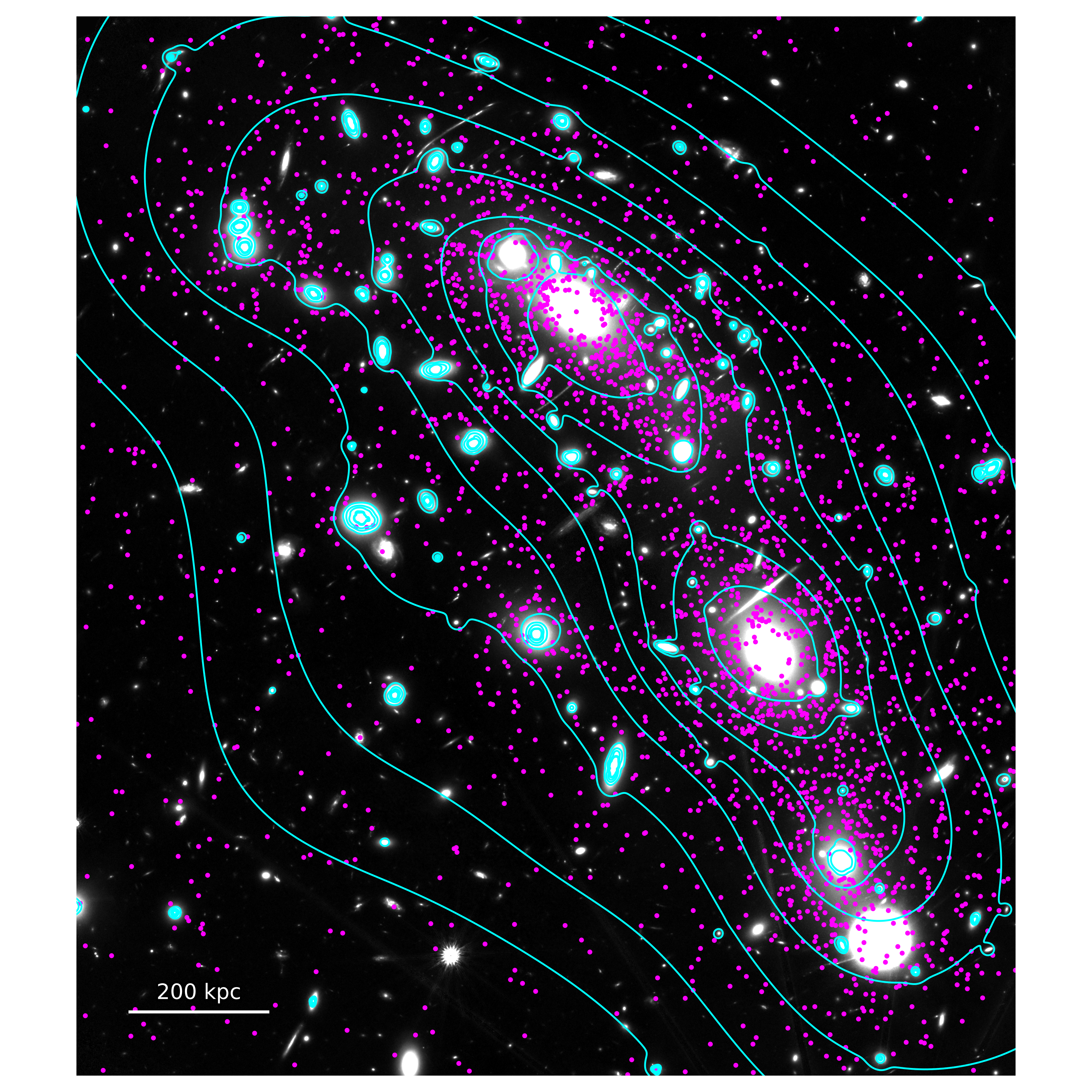}
\caption{\textit{Left}: The MACS0416 full mosaic field with north at the top and east to the left. The red box outlines the central cluster and the region used in this analysis. As described in the text, the outer region of the mosaic image with shorter total exposures and higher sky noise was not used, except for increasing our stellar PSF sample. 
\textit{Right}: Pink dots show the locations of the 2,971 GC candidates 
for MACS0416 found in this study. Cyan contours show the projected mass density from the lensing model of \citet{diego2023jwst}.
}
\label{fig:mosaic}
\end{figure*}


\section{Data and Photometry} \label{sec:data}

The PEARLS program observed MACS0416 in eight filters spanning 0.9 to 4.4 $\micron$  $(F090W,$ $F115W,$ $F150W,$ $F200W,$ $F277W,$ $F356W,$ $F410M,$  $F444W)$ in three epochs between October 2022 and February 2023 \citep{windhorst2023jwst}. With total integration times ranging from 2,920 to 3,779 seconds per filter, the observations reach 5$\sigma$ depths of 28.45 to 29.9 AB mag for point sources, sufficient to resolve GCs against the cluster background or its ICL \citep[\eg][]{diego2023jwst}. The exposures were initially all calibrated with the \textit{JWST} pipeline\footnote{\url{https://github.com/spacetelescope/jwst}} version 1.13.4 \citep{Bushouse2024}, using calibration files from version 1230 of the Calibration Reference Data System (CRDS)\footnote{\url{https://jwst-crds.stsci.edu}}, with special attention paid to removal of 1/f noise and other low-level artifacts \citep[further details in][]{windhorst2023jwst}. All the calibrated exposures were subsequently aligned and drizzled onto mosaics at 0.030 arcsec/pix, following the techniques first described by \cite{koekemoer2011}, updated to use the \textit{JWST}. The mosaics extend over several arcminutes, centered on the cluster, with additional irregular parallel coverage. The central region thoroughly covers the cluster core  ($1.62^{\prime} \times 1.83^{\prime}$ which corresponds to $\sim520 \times 580$ kpc at $z=0.397$ as indicated in Fig.~\ref{fig:mosaic}), and is included in all three epochs. In this region, the short-wavelength NIRCam filters reach cumulative exposure times of 11,338\,s in $F090W$ and $F115W$, and 8,761\,s in $F150W$ and $F200W$, with each filter composed of 12 dithered exposures (\(n_{\mathrm{exp}} = 12\)) (\citet{windhorst2023jwst}; for details about these observations, see also \citet{yan2023webb} for a detailed description of the cluster and data.)  

Our analysis will focus on this central region, as shown in Figure \ref{fig:mosaic}. However, due to the scarcity of foreground stars in the central area, the outer regions were used to provide a sample of unresolved (star-like) objects, which were later used to build point spread function (PSF) models. 

We have chosen to omit the NIRCam long-wavelength (LW) data from our GC analysis. While filters such as $F277W$, $F356W$, $F410M,$ and $F444W$ provide useful long-wavelength leverage for tasks like AGN identification and spectral energy distribution (SED) fitting, they are less suitable for primary GC color--magnitude analyses at this redshift due to their broader PSFs, lower sensitivity, and limited diagnostic power for metallicity.  
For example, to achieve $S/N = 10$ in a 10 ks exposure for point sources, F410M requires $\sim$22.1\,$\mu$Jy (AB~$\approx$~28.0 mag) and F444W requires $\sim$17.4\,$\mu$Jy (AB~$\approx$~28.3 mag), compared to just 8.0\,$\mu$Jy (AB~$\approx$~29.1 mag) for F200W.\footnote{Sensitivity values from STScI NIRCam documentation: \url{https://jwst-docs.stsci.edu/jwst-near-infrared-camera/nircam-performance/nircam-sensitivity}}  
Nevertheless, we emphasize that the LW imaging remains valuable for complementary science. In particular, the additional color leverage at $2{-}5\,\micron$ will help define the locus of GC candidates in future work, refine stellar population diagnostics, and constrain the separation of genuine GCs from contaminants. In the present study, however, the broader PSFs and comparatively lower signal-to-noise for faint, compact sources in the LW filters made them less effective for GC selection, and so they were not adopted as a formal criterion.

\subsection{Photometric Methods with \textsc{DAOPHOT}}

Photometry was performed with the tools in the package \textsc{DAOPHOT} in {\sc IRAF} \citep{stetson1987DAOPHOT}. Six images $(F090W,$ $F115W,$ $F150W,$ $F200W,$ $F277W,$ $F356W,$ $F444W)$ were stacked to produce an ultradeep image for object finding. The \texttt{daofind} task produced a list of 25,720 candidate sources within the 1.62\arcm $\times$ 1.83\arcm central region, later refined through iterative analysis, as described below. For each of the SW filters ($F090W$–$F200W$), we adopted a detection threshold of $4\sigma$ above the local background, used a Gaussian convolution kernel matched to the PSF width, and required detections to satisfy \texttt{sharplo}=0.2, \texttt{sharphi}=1.0, \texttt{roundlo}=$-1.0$, and \texttt{roundhi}=1.0. Prior to running \texttt{daofind}, we subtracted a smoothed background map (mesh size $29\times29$ px) to reduce the contribution from diffuse intracluster light (ICL) and improve uniformity across the field. With the candidate list, the \texttt{phot} task performed preliminary aperture photometry, with aperture radii of 3 pixels (0.090\arcs) for flux measurement in the SW camera. 
Iterative sigma-clipping was used to calculate the mean sky-values in the sky annulus, with a maximum of 50 iterations.  Sigma-clipping is needed to remove the contaminating effects of faint objects located in the sky annulus (i.e., crowding effects). 

Direct measurements of unsaturated stars in the mosaics give PSF sizes of FWHM = 1.8 pixel $(F090W)$, 1.9 pixel $(F115W)$, 2.1 pixel $(F150W)$, and 2.4 pixel $(F200W)$. These PSF FWHM values were also used to set the parameters of the \texttt{daofind} convolution kernel in each filter. We verified that PSFs constructed independently from stars in the central cluster region and from the parallel field produce consistent photometric results, confirming that any PSF broadening due to overlapping rotated pointings in the core is negligible. The \emph{psf} task constructed a variable PSF model through 1) a selection of 7 isolated stars across the field; 2) Iterative PSF fitting with 1st-order spatial variations; and 3) Neighbor subtraction using the \emph{allstar} task with a 3-pixel (0.090 arcsec) fitting radius. In the Milky Way, the half-light diameter of the GC ranges from $\sim$1 to 10 pc, with a median near 5 pc \citep{kundu1999globular, harris2010diamonds}. Consequently, individual GCs in MACS0416 appear as unresolved (point-like) sources.

Next, the photometry output list from each filter was matched to within 1 pixel, accepting objects that were only matched in \emph{all four filters} ($F090W,$ $F115W,$ $F150W,$ and $F200W$). 
Matching across all four filters is helpful for eliminating a significant number of false detections, such as residual (unfiltered) cosmic rays, artifacts, noise clumps, and similar issues. This multi-filter detection requirement provides a highly reliable GC candidate sample while introducing only minimal color-dependent bias. The 5$\sigma$ point-source depths of the SW mosaics are 29.9, 29.6, 29.3, and 28.9~AB~mag for $F090W$, $F115W$, $F150W$, and $F200W$, respectively \citep{windhorst2023jwst}. At $z = 0.397$, the intrinsic spectral energy distributions of GCs vary by $\lesssim$0.2~mag across these filters for the full range of metallicities \citep{hartman2025sed}, implying that virtually all GCs bright enough to be detected in $F200W$ are also detected in $F090W$. The only noticeable completeness variation occurs near the faint limit, where the detection boundary becomes slightly color-dependent. However, the bright portion of the CMD, used for all subsequent analyses, is fully complete and unaffected by this small bias.

\begin{figure}[t]    
\centering{
  \includegraphics[width=1\hsize]{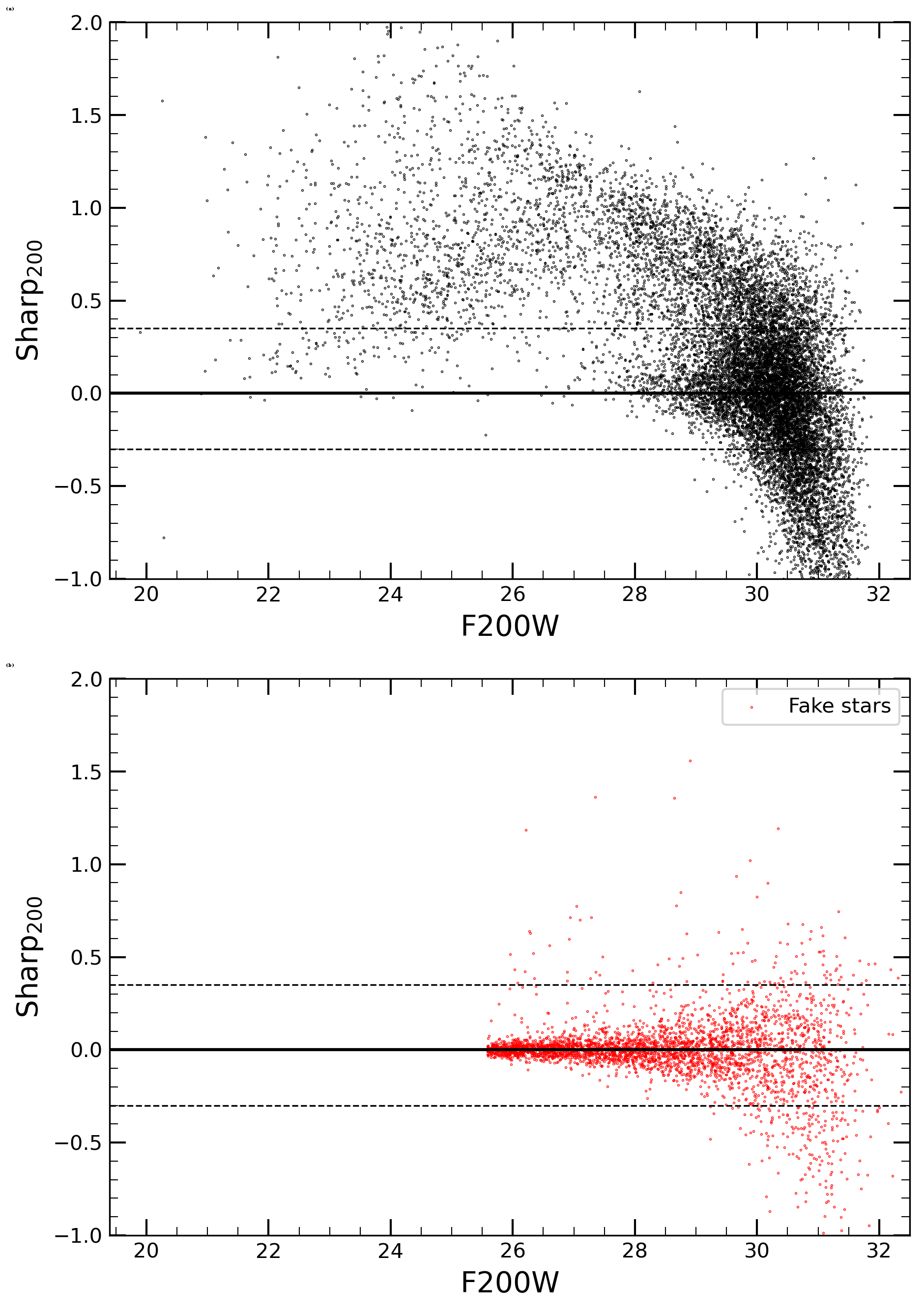}
  \caption{\textit{Upper panel}: \textsc{DAOPHOT} \emph{sharp} parameter versus $F200W$ magnitude for all detected sources. The dashed horizontal lines mark the adopted selection range ($-0.3 < \textit{sharp}_{200} < 0.35$), outside of which objects are excluded as non-stellar or poorly fit by the PSF. 
\textit{Lower panel}: Results from artificial star tests, showing that genuine point sources occupy the same sharpness range and thus motivate the adopted boundaries. A bright-end cut at $F200W < 26$ is also applied, consistent with the luminosity limits of confirmed GCs at $z\sim0.4$ \citep{harris2023jwst, harris2024jwst}.}}
\label{fig:sharp_plot}
  
\end{figure}

To further reduce sample contamination, we followed methods from \citet{harris2009globular} using the \textsc{DAOPHOT} \emph{sharp} parameter to reject objects that did not exhibit stellar profiles consistent with the PSF. As shown in Figure~\ref{fig:sharp_plot}, rejected sources fall outside the restricted range between \emph{sharp} boundaries of $-0.3$ and $+0.35$. These limits were determined by inserting artificial stars across the observed magnitude range and identifying the locus of genuine point sources (Fig.~\ref{fig:sharp_plot}, lower panel). Objects outside this range are predominantly extended galaxies or noise peaks and were excluded from the GC sample. 

We also applied a bright-end magnitude cut at $m_{F200W} < 26$, following recent JWST studies of cluster GCs at similar redshifts \citep{harris2023jwst, harris2024jwst}, which show that the most luminous bona fide GCs rarely exceed this brightness. Sources brighter than this threshold are retained in a separate catalog for analysis in Section~6.1, where we examine their properties in the context of ultra-compact dwarfs (UCDs) and stripped nuclei.

\subsection{Assessing and Correcting Background Contamination}

Having established our source detection and photometric measurements, we next assess the contribution of background and foreground contaminants to the GC candidate sample.
To estimate the contribution of point-like contaminants—primarily foreground stars and unresolved background galaxies—to our GC candidate sample, we applied a spatially uniform dilution test. This method assumes that contaminants are approximately uniformly distributed across the field, in contrast to the centrally concentrated distribution expected for true cluster GCs. 

We divided the image into a fixed grid of equal-area spatial cells (see Appendix~\ref{sec:appendix_contam}). Within each cell, we randomly removed up to 10 sources. This number was chosen based on the typical source counts in the outermost (off-cluster) regions, where the GC surface density flattens and background contamination dominates. In sparsely populated cells containing 10 or fewer sources, all were removed; in more well populated cells, exactly 10 were randomly excluded. This conservative approach yields a statistical background correction without requiring photometric or radial cuts, and avoids introducing biases tied to GC color or magnitude.

By performing the dilution locally within each cell, we reduce sensitivity to possible small-scale clustering or non-uniformities in the contaminant population. Although this method does not explicitly account for contaminants that may correlate with lensing structures or Galactic foreground variations, it provides a useful upper limit on the contamination fraction under the assumption of a spatially uniform background. Importantly, the same GC candidate selection criteria described above (multi-filter detection, sharpness and roundness cuts, magnitude limits, and PSF-based classification) were applied uniformly to both the central cluster field and the outer pointings. Thus, the estimated $\sim$10 sources per cell reflects the residual background contamination \emph{after} GC candidate selection, rather than raw detections.

We note that some genuine GCs, especially intergalactic or intracluster candidates in the outskirts of the cluster, may be inadvertently subtracted by this method. A more refined correction would involve isolating sources along the GC color sequence and computing spatial densities within a restricted color–magnitude range. However, for the present analysis, this dilution test offers a reasonable first-order correction. In total, 549 sources were removed from the catalog, providing a statistical estimate of the background contribution.

We emphasize that the primary goal of this work is to trace the extended spatial distribution of GCs across the cluster. While more aggressive modeling of galaxy light profiles (e.g., \citealt{cho+2016,Lee2022}) could yield additional detections closer to the bright galaxy centers, such a step would primarily aid in recovering the \emph{brightest} GCs projected against the inner halos, where the local surface brightness gradients are steepest.  Galaxy-subtracted images will also be better for visualizing the GC distribution in the central regions. However, the completeness limit of the photometry will be considerably brighter in these inner regions because the limit is determined by the local sky noise rather than simply the diffuse light level. Thus for purposes like the GC luminosity function analysis (see below) the outer halo regions where much deeper point-source detections can be made remain the most important. This behavior is discussed in more detail in \citet{harris2024speagle}, whose analysis emphasizes the same trade-off between depth and crowding effects in dense cluster fields.

\begin{figure}[t]    
\centering{
  \includegraphics[width=1\hsize]{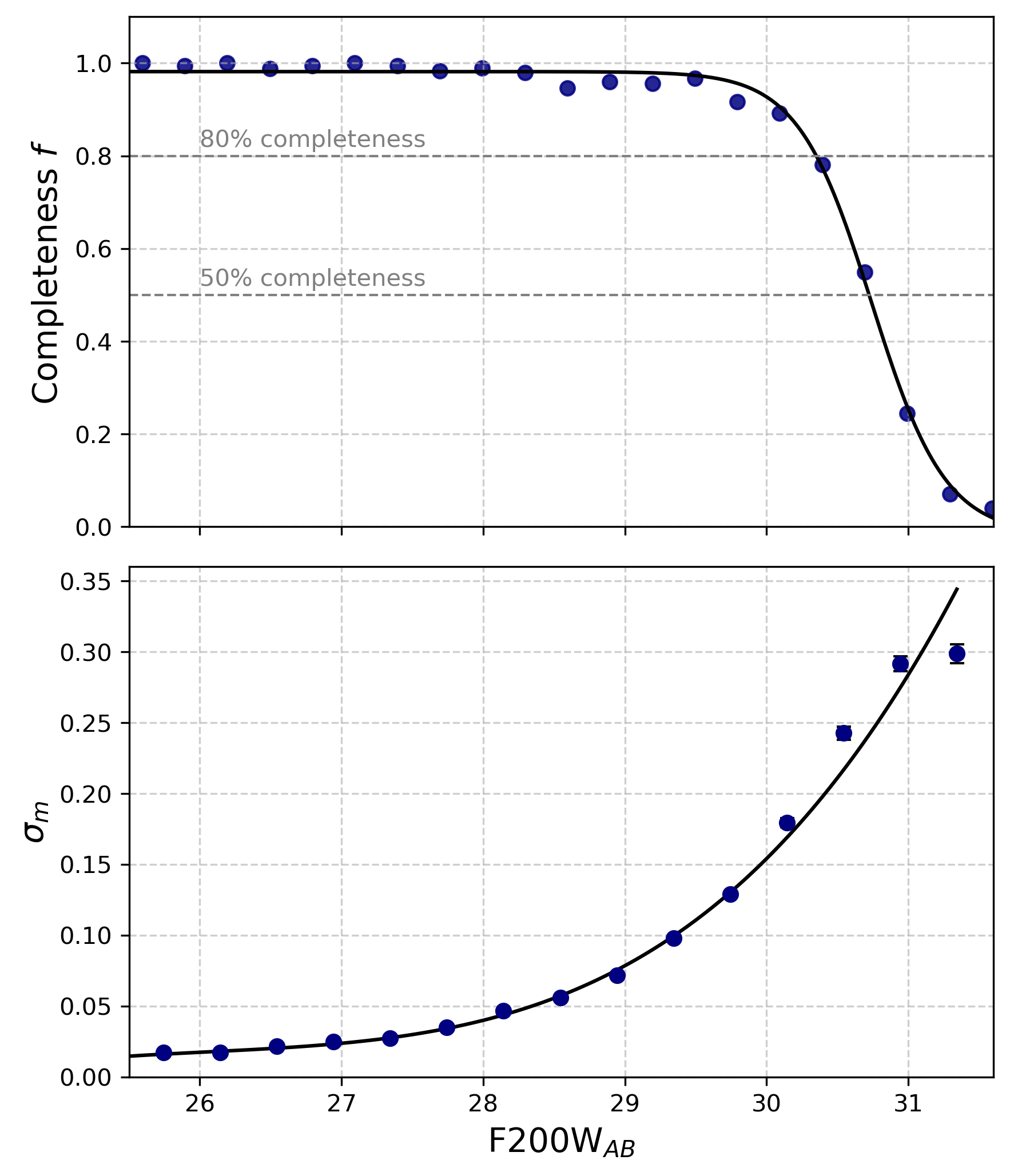}
  \caption{\textit{Upper panel}: Photometric recovery completeness measured from artificial star tests. The recovery fraction $f$ is plotted in 0.3-magnitude bins. The modified hyperbolic tangent function given in the text is shown as the solid line. \textit{Lower panel}: Internal measurement uncertainty $\sigma$ versus magnitude,
    as determined from the artificial-star tests.}\label{fig:stats_plot}
  }
  \end{figure}
\begin{figure}    
\centering{
  \includegraphics[width=1\hsize]{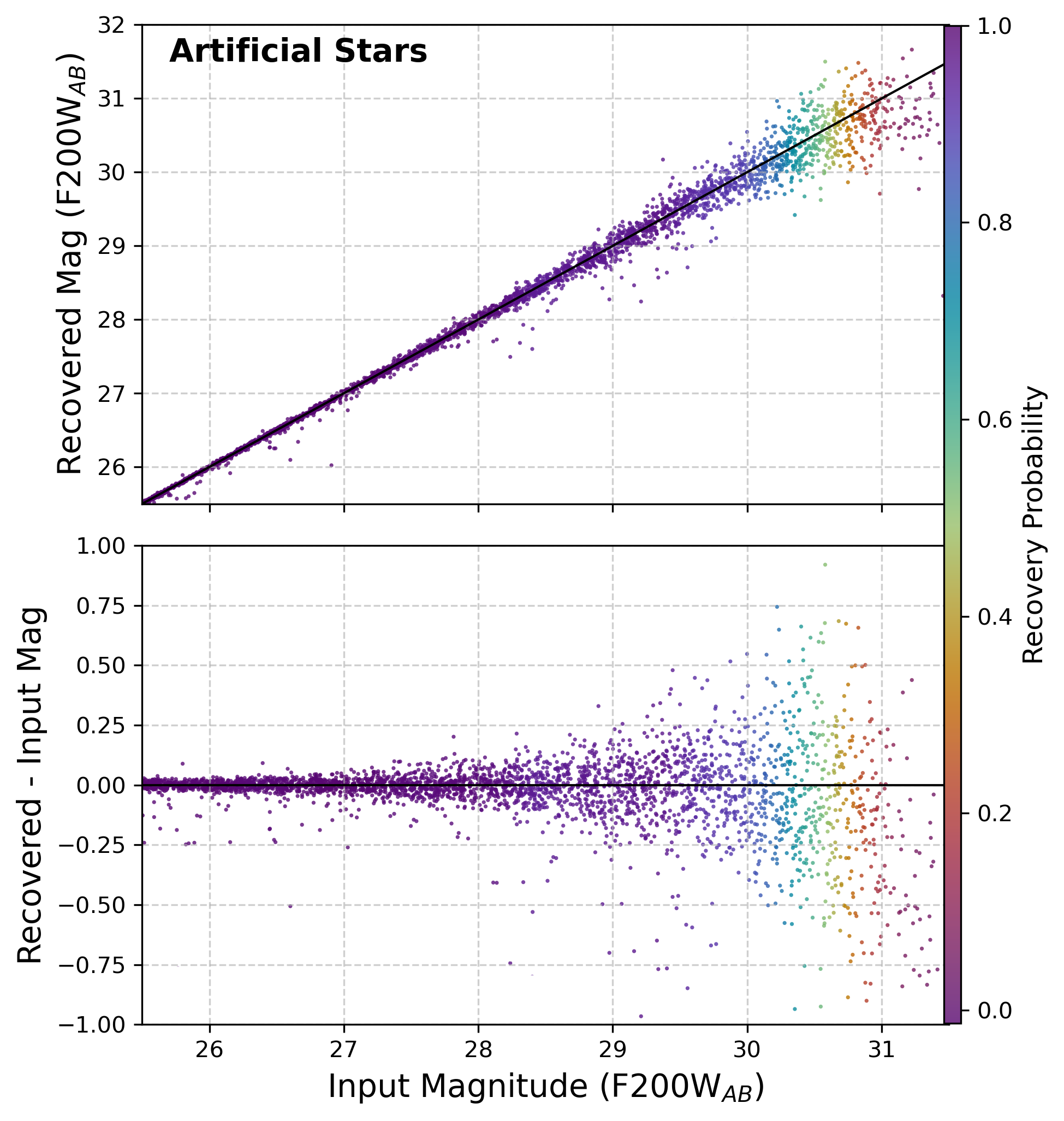}
  
  \caption{\textit{Upper panel}: Measured recovery magnitude in $F200W$ of the artificial stars vs. input magnitude and color-coded by their recovery probability. The solid line shows the 1:1 relation. \textit{Lower panel}: Residuals (recovered -- input magnitude) as a function of input magnitude. At brighter magnitudes, stars are accurately recovered, while systematic deviations and increased scatter appear at fainter magnitudes near the detection limit.  }\label{fig:art_stars}
  }
  \end{figure}
\begin{figure*}
    \centering
    \includegraphics[width=.9\textwidth,height=\textheight,keepaspectratio]{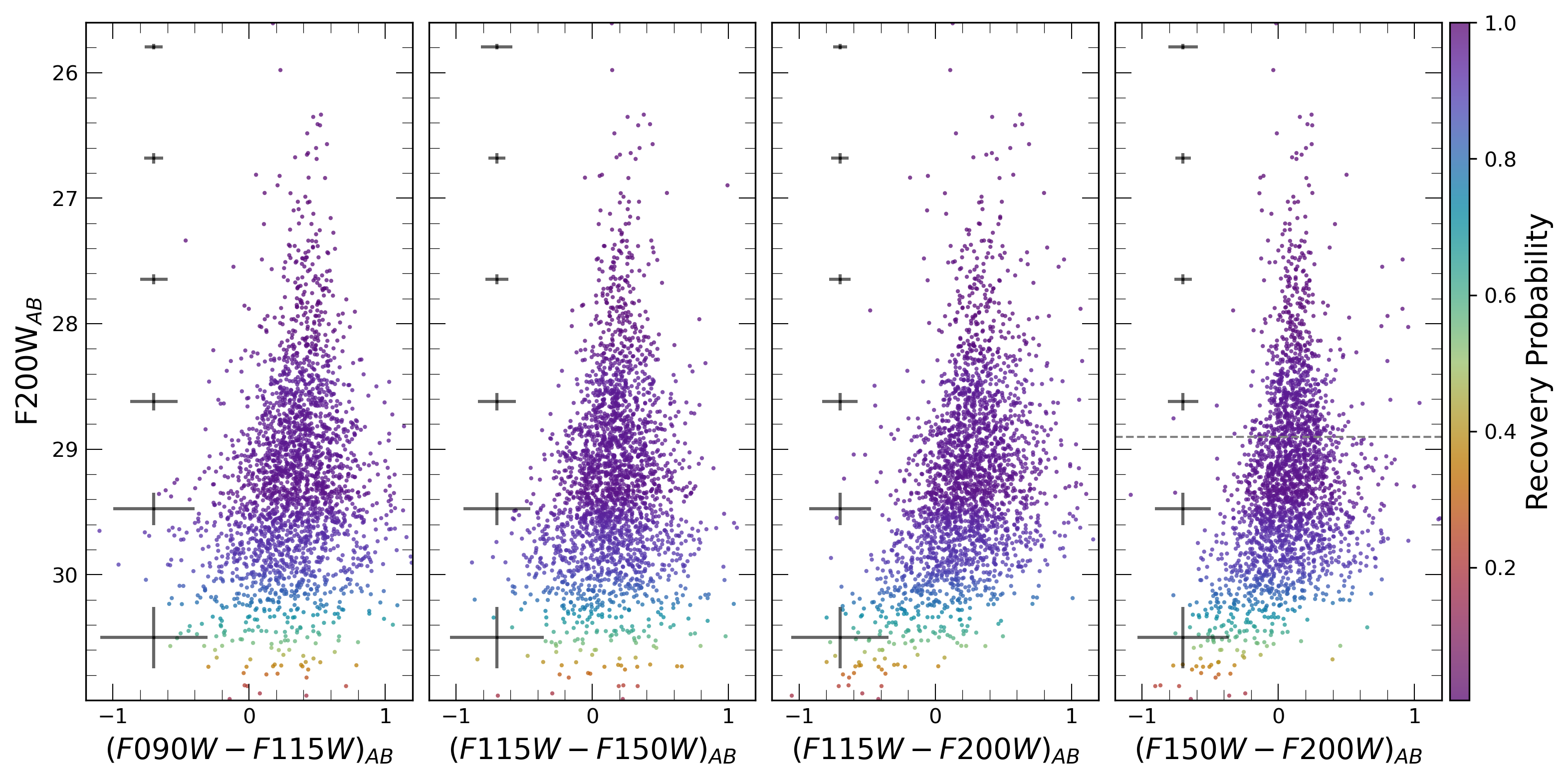}
    \caption{CMDs of unresolved point-like sources in MACS0416, showing intrinsic color versus apparent $F200W$ magnitude for four filter combinations. Each point is color-coded by the photometric recovery probability derived from artificial star tests, interpolated as a function of $F200W$ magnitude. Importantly, the completeness values shown here reflect the joint detection probability across all four NIRCam filters ($F090W$, $F115W$, $F150W$, and $F200W$). Because inclusion in the CMD requires successful detection in all bands, the joint completeness decreases more rapidly at faint magnitudes than in any individual filter. The color bar ranges from 0 to 1.0 and illustrates this combined recovery probability. Error bars on the left side of each panel show the median photometric uncertainty in color and magnitude, computed in integer bins of $F200W$ magnitude. These bars highlight the increase in color uncertainty toward the faint end. In the last panel, the horizontal dashed line represents the 5$\sigma$ completeness limit.} No K-corrections have been applied.
    \label{fig:CMD}
\end{figure*}

\bigskip

\section{Completeness test and recovery probability} \label{sec:completeness}

One of the key applications of \textsc{DAOPHOT} is the {\sc addstar} function to perform artificial star tests to evaluate the recovery probability and completeness of a GC data sample. 

Prior to the completeness simulations, we characterized the point-spread function (PSF) across the NIRCam mosaics using isolated stars in both the outer single-field regions and the central overlapping area. The empirical FWHMs measured in the outer fields are consistent with the theoretical JWST/NIRCam values—approximately 0.067$^{\prime\prime}$ (F090W), 0.074$^{\prime\prime}$ (F115W), 0.086$^{\prime\prime}$ (F150W), and 0.105$^{\prime\prime}$ (F200W)—to within 5--10\%. As expected, the stacked exposures in the cluster core, which combine multiple rotated pointings, produce slightly broader and more symmetric PSFs, with median FWHM increases of $\approx$3--5\%. This small variation has a negligible impact on the derived photometry or completeness, since all fluxes are measured with PSF-fitting routines and aperture corrections are based on the same empirical PSFs used for the artificial-star tests.

In this test, a set of approximately 5,000 artificial stars with magnitudes ranging from
$26.5-31.5$ AB mag was inserted into each of the six images mentioned above. The artificial stars used in each of the four images all share the same input x and y coordinates and magnitudes. These artificial stars are then processed using the same photometric procedures outlined in the above section, again, using only the $F090W,$ $F115W,$ $F150W,$ and $F200W$ images for our GC analysis. The recovery probability refers to the likelihood that an artificial star will be correctly identified and measured and is assessed by comparing the input properties of the artificial stars to the measured properties in the output catalog. The top panel in Figure \ref{fig:stats_plot} shows the recovery fraction $f$ of the artificial star sample on the $F200W$ image that have also passed the \emph{sharp} test. The recovery fraction is determined by calculating the number of artificial stars that are recovered in each given 0.3 mag bin, divided by the number of input stars in that bin. This method helps to assess the depth to which the data can reliably detect GCs. The curve in the top panel of Figure \ref{fig:stats_plot} can be fit with a scaled and shifted modified hyperbolic tangent function \citep{harris2023jwst} of the form:
\begin{equation}
f(m) = 0.437 - 0.547 \cdot \tanh\left[\beta (m - m_1)\right]
\end{equation}
\n where parameter $\beta$ = 2.26 controls the steepness of the transition and $m_{1}$ = 30.63 mag corresponds to the magnitude at 50\% completeness, yielding an 80\% completeness level of $\sim$30.36 mag. The complete recovery of $F200W_{80\%}$ and $F200W_{50\%}$ is shown in Fig. \ref{fig:stats_plot} as gray dashed lines. This functional form is mathematically equivalent to a rescaled logistic function but provides a slightly sharper and more symmetric transition, which better reflects the rapid decline in detection probability seen in our data. The model ensures physically meaningful behavior over the full magnitude range, with completeness smoothly declining from 1 at the bright end to 0 at the faintest catalog fluxes.
The resulting recovery fraction is also shown in the color bars of Figures \ref{fig:art_stars} and \ref{fig:CMD}. The lower panel in Figure \ref{fig:stats_plot} shows how the associated photometric measurement uncertainty $\sigma$ increases towards fainter fluxes.

Figure~\ref{fig:art_stars} shows a general trend in which recovered magnitudes closely follow input magnitudes, especially for brighter objects, indicating a high recovery probability (\gtrsim95\%) for sources brighter than 29.64 mag in $F200W$. The probability of recovery drops significantly for fainter objects due to lower signal-to-noise and reduced detection efficiency. Artificial stars were injected to probe the full dynamic range, though we did not include stars brighter than 26.5 or fainter than 31.5 due to the magnitude range expected of GCs. 

After assessing photometry and completeness, we construct CMDs of point-like sources in MACS0416, as shown in Figure~\ref{fig:CMD}. Each panel plots $F200W$ magnitude against a different color index, using a matched sample of objects detected in all four filters ($F090W$, $F115W$, $F150W$, and $F200W$). Recovery probability, derived from the artificial star tests, is color-coded for each source. The color bar reflects joint completeness, defined as the fraction of artificial stars recovered in all four filters simultaneously. As such, it declines more rapidly than the completeness in any individual filter. 

The CMDs extend to $F200W$ $\sim$ 31 mag in individual filters; but any statistical analysis should take into account the more stringent joint completeness that governs the CMD sample.
The artificial stars injected for completeness testing were modeled as point sources with flat spectra in the AB system (i.e., constant $f_{\nu}$), corresponding to zero color across the NIRCam filters. This implicitly represents the average GC spectral energy distribution (SED) near the rest-frame optical, as seen in recent JWST studies of cluster GCs \citep{hartman2025sed, harris2024jwst}. Because completeness was evaluated independently in each filter, the combined four-band detection requirement introduces only a mild color dependence near the faint limit, where the bluer filters reach slightly shallower depths. However, for the bright regime ($m_{F200W} \lesssim 29$ mag) that dominates our GCLF and color analyses, the completeness is effectively color-independent. The adopted completeness corrections, therefore, remain robust for the GC population analyzed here.
This implies that any statistical analysis (e.g., LFs or GC color) should account for incompleteness around and below this flux level. 
 
\section{Color Distribution and Bimodality Tests} \label{sec:colors}

A key diagnostic of globular cluster populations is the shape of their color distribution function (CDF), which reflects the underlying age and metallicity structure. In massive galaxies, photometric studies using \textit{HST} have consistently revealed bimodal GC color distributions—typically interpreted as metal-poor (blue) and metal-rich (red) subpopulations \citep[\eg][]{blakeslee1999globular, lee2010detection, peng2011hst, beasley2020globular, harris2023photometric}. These distributions are often well fit by double-Gaussian models and are thought to trace distinct formation channels despite similar internal properties \citep[\eg][]{ashman1992formation, forbes1997origin, cote1998formation, muratov2010modeling, harris2023jwst}.

The near-infrared coverage of \textit{JWST}/NIRCam complicates this classical picture. Although highly effective for detecting faint, distant ($z > 1$) objects \citep[\eg][]{mowla2022sparkler, lee2022detection}, the SW camera color indices, such as $F115W-F200W$ and $F150W-F200W$, span narrower wavelength baselines in the near-IR, where metallicity sensitivity is reduced. As a result, GC CDFs in JWST studies often appear unimodal, or show only weak structure \citep{berkheimer2024jwst, harris2024jwst}. In light of this, we use the term “multiple components” rather than “bimodality” when interpreting our color distributions in MACS0416.

To evaluate whether a multiple-component model provides a better description than a single Gaussian, we applied the KMM algorithm \citep{ashman1994detecting} and Gaussian Mixture Modeling (GMM) to the $F115W-F200W$ and $F150W-F200W$ color distributions. The KMM method uses a likelihood ratio test (LRT):

\begin{equation}
\Lambda = 2 \left( \log \mathcal{L}_2 - \log \mathcal{L}_1 \right),
\end{equation}

\noindent to compare the goodness of fit for bimodal ($\mathcal{L}_2$) versus unimodal ($\mathcal{L}_1$) Gaussian models. The test statistic

\begin{equation}
P_{\mathrm{KMM}} = 1 - F_{\chi^2_{k}}(\Lambda)
\end{equation}

\noindent assumes that both components share the same variance, and yields the probability of incorrectly rejecting a unimodal model \citep{ashman1994detecting, chies2012optical}. A high $\Lambda$ and low $P_{\mathrm{KMM}}$ indicate statistically significant preference for a multi-component model.

We chose not to include color indices involving $F090W$ (e.g., $F090W-F200W$) in our multi-component analysis due to significantly larger photometric uncertainties in the F090W band, especially at the faint end. These increased errors would artificially broaden the color distribution and diminish our sensitivity to potential substructure. 

To test the stability of our results, we applied KMM to samples limited by photometric completeness thresholds of 95\% and 97\%, corresponding to magnitude limits of 29.64 and 28.44 mag, respectively. We also used GMM fitting with two components to examine the Bayesian Information Criterion (BIC), Akaike Information Criterion (AIC), and component separation $\Delta \mu$ (mean color difference). The results of both tests are summarized in Table~\ref{tab:KMMresults}.

For the $F115W-F200W$ color index, the KMM results indicate strong statistical evidence for multiple components at 95\% and 97\% completeness ($P_{\mathrm{KMM}} \ll 0.01$), with modest mean separations of $\Delta\mu \sim 0.02$mag. This trend likely reflects increased photometric scatter at fainter magnitudes and is consistent with prior findings of weak bimodality in the near-infrared \citep[\eg][]{Peng2006, Cantiello2007, chies2012optical,cho+2016}.

For $F150W-F200W$, the 95\% sample returns a large $\Lambda = 193$ with $P_{\mathrm{KMM}} \approx 0$, indicating a statistically significant preference for a double-Gaussian model over a single Gaussian model, despite the small component separation of $\Delta\mu = 0.006$~mag. In contrast, the 97\% sample yields $\Lambda = 2.7$ and $P_{\mathrm{KMM}} = 0.44$, providing no significant evidence for multiple components. \deleted{Similarly, the 99\% sample gives $\Lambda = 1.8$ and $P_{\mathrm{KMM}} = 0.62$, reinforcing the lack of bimodality at high completeness thresholds.} This may result from poorer statistics at the high completeness, or from enhanced photometric covariance between these two closely spaced bands \citep[see also][]{choksi2018formation, Pozzetti2019}. 

A future study will use Hubble Frontier Fields imaging in a follow-up study to investigate the optical colors of the brightest GCs in MACS0416.

\begin{figure*}    
\centering{
\includegraphics[width=.9\textwidth,height=\textheight,keepaspectratio]
  {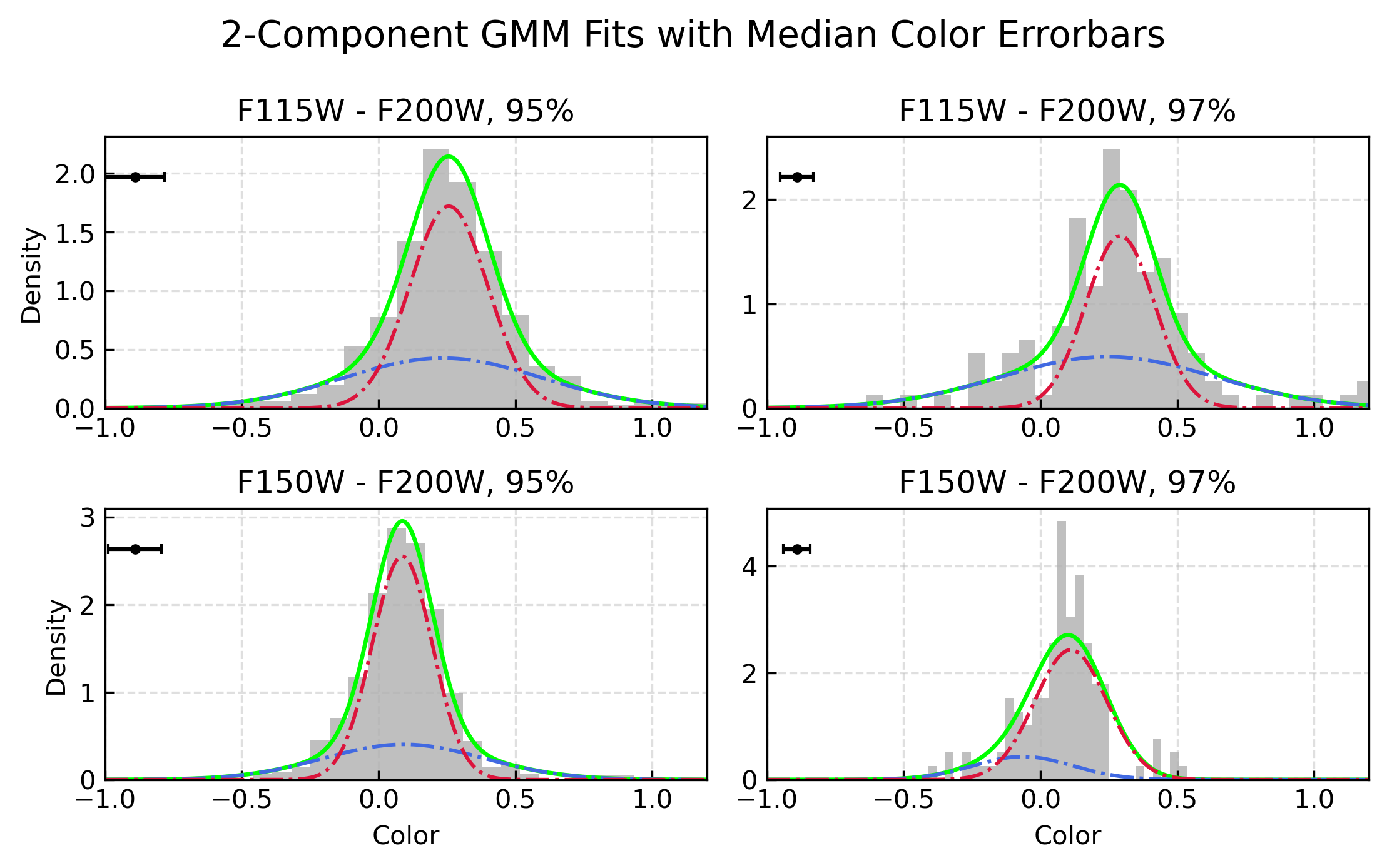}
  \caption{Two-component GMM fits to the $F115W-F200W$ (top) and $F150W-F200W$ (bottom) color distributions at the 95\% and 97\%completeness levels. Histograms show our MACS0416 GC data, with individual Gaussian components (dashed lines) and total fits (solid lines) overlaid. Error bars in the upper-left corners display the photometric color error in the displayed completeness level.}\label{fig:hist_grid}
  }
\end{figure*}

\begin{table}[ht]
\centering
\caption{KMM outputs for the color distributions.}
\begin{tabular}{ccccc}
\hline\hline
\% compl & color index & $\Lambda$ & $\Delta\mu$ (mag) & $P_{\mathrm{KMM}}$ \\
\midrule
95\% & $F115W-F200W$ & 153.24 & 0.022 & 0.000 \\
97\% & $F115W-F200W$ & 22.31  & 0.045 & 0.000 \\
\midrule
95\% & $F150W-F200W$ & 192.81 & 0.006 & 0.000 \\
97\% & $F150W-F200W$ & 2.73   & 0.175 & 0.440 \\
\bottomrule
\end{tabular}
\label{tab:KMMresults}
\end{table}

\section{The GCLF Turnover} \label{sec:gclf}

The GCLF is well known to exhibit a near-universal log-normal form in the local Universe for massive galaxies, with a characteristic turnover at an absolute magnitude of $M_V \simeq -7.4$ \citep[e.g.,][]{rejkuba2012globular, harris2014globular}.
At the redshift of MACS0416, this corresponds to an apparent turnover magnitude of $V \sim 34.3$, assuming no evolution and using a distance modulus of $(m\!-\!M)_0 = 41.73$ mag. However, empirical and theoretical models predict that the turnover should appear at somewhat lower luminosity at earlier cosmic times due to simple stellar evolution and the progressive dynamical destruction of low-mass clusters. For example, the EMP-Pathfinder models of \citet{reina2022introducing} suggest a shift of $\sim0.3-0.4$~mag fainter in $M_V$ at a lookback time of 4.4~Gyr. This evolutionary effect is distinct from the mass loss experienced by individual GCs, which is relatively modest for high-mass clusters near the bright end of the GCLF. As shown by \citet{choksi2018formation}, the amount of mass lost depends strongly on initial mass, and the most massive GCs retain much of their mass over time.  All clusters, however, will show a decrease in luminosity with time due to stellar evolution and the progressive loss of high-mass stars.

Detecting the turnover becomes increasingly difficult at intermediate redshifts due to cosmological surface brightness dimming, rising photometric uncertainties, and filter bandpass shifts, all of which reduce the effective photometric depth and completeness. While these effects are more severe in optical imaging, near-infrared observations (as employed here) benefit from more favorable $K$-corrections, which can partially offset the flux loss by shifting rest-frame optical light into redder, more sensitive bands \citep[\eg][]{Peng2009, AlamoMartinez2013, Harris2017}. Although younger GC systems may contain brighter, less-evolved clusters on average \citep[\eg][]{Kruijssen2015, ElBadry2019, Choksi2018}, this potential luminosity enhancement does not compensate for the steep completeness falloff at faint magnitudes as shown in Figure \ref{fig:stats_plot}. Moreover, the requirement of multi-band detection (as in our CMD-based analysis) further suppresses the recovery of faint GCs, rendering direct measurement of the turnover well beyond the reach of the current dataset. Nevertheless, the observed portion of the LF remains sufficient to assess its overall shape and consistency with a log-normal shape at the bright end.

Despite these challenges, the GCLF remains a valuable diagnostic of globular cluster populations, particularly when survey depth and completeness corrections are properly characterized.

\subsection{K-Corrections} \label{sec:kcorr}

The cosmological $K$-correction has long been a critical tool in galaxy studies, enabling accurate comparisons of rest-frame luminosities across redshift \citep[\eg][]{mannucci2001near, blanton2007k, fielder2023empirically}. With \textit{JWST}'s ability to observe extragalactic GCs at much greater distances than previously possible, it is necessary to apply appropriate $K$-corrections when interpreting apparent magnitudes.

For this analysis, we adopt the GC $K$-correction framework described in \citet[\eg][]{reina2024rescuer}. At the redshift of MACS0416 ($z = 0.397$), the corresponding lookback time is approximately 4.38 Gyr. Assuming a typical GC age of 7 Gyr at that lookback time, corresponding to 11.4 Gyr in the local Universe \citep{vandenberg+2013,forbes_bridges2010,dotter+2011}  and a mean metallicity of [M/H] = $-1$ \citep[\eg][]{Peng2006, harris2010diamonds, harris2013catalog}, we compute $K$-corrections for each \textit{JWST}/NIRCam filter using the online RESCUER tool\footnote{\url{https://rescuer.streamlit.app/}}. These values are not highly sensitive to modest variations in assumed age or metallicity.

Table~\ref{tab:kcorr} summarizes the resulting corrections. The quantity $A_{\lambda}$ denotes the foreground Galactic extinction in each filter, calculated from the \citet{schlafly2011measuring} dust maps and scaled using the extinction law of \citet{fitzpatrick1999correcting}. The filter-dependent $K$-corrections, $K_{\lambda}$, are expressed in magnitudes and reflect the shift between rest-frame and observed-frame photometry for GCs at $z = 0.397$ under the adopted parameters.
\begin{table}
\centering{
\caption{Extinctions and K-corrections.} \label{tab:kcorr}
\begin{tabular}{ccc}
  \hline \hline
Filter & $A_{\lambda}$ (mag) & $K_{\lambda}^{\star}$ (mag) \\
   \midrule
   F090W & 0.037 & $-0.063$ \\
   F115W & 0.032 & $-0.244$ \\
   F150W & 0.020 & $-0.210$ \\
   F200W & 0.013 &$-0.465$ \\ \hline
 
\multicolumn{2}{l}{${\star}$ Assumes $\tau=7$ Gyr, [M/H] = -1.} \\

\end{tabular}}
\end{table}

\subsection{GCLF fitting} \label{sec:LFfitting}


\begin{figure}    
  \centering
  \includegraphics[width=1\hsize]{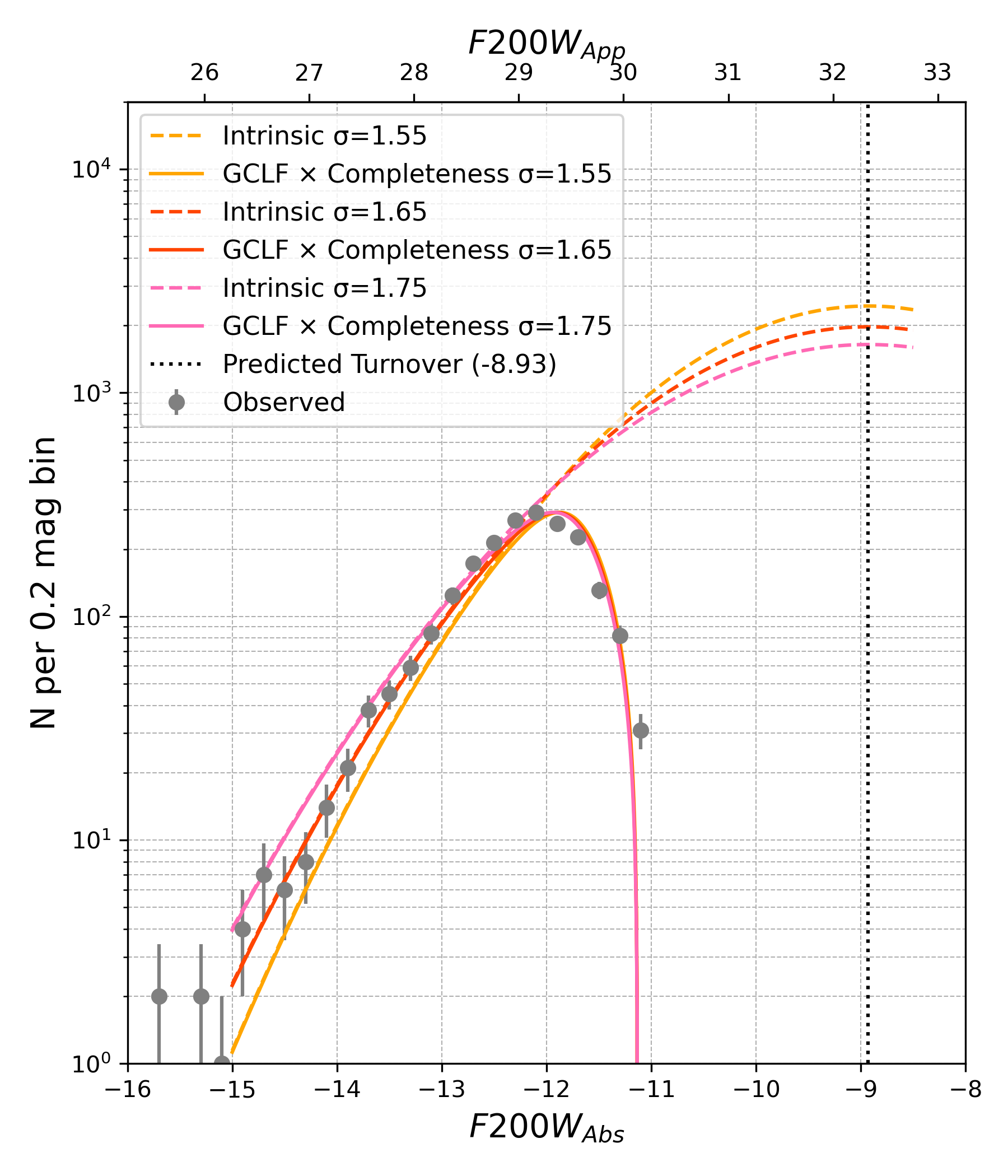}
  \caption{K-corrected GCLF based on F200W photometry. Gray points show the observed number counts per 0.2 mag bin. Overlaid are intrinsic log-normal GCLF models (dashed lines) with $\sigma = 1.55$, $1.65$, and $1.75$, and their completeness-modified counterparts (solid lines) based on the best-fit detection function (Equation~1). The model curves have been scaled to match the observed peak. The vertical dashed line marks the estimated turnover magnitude after k-correction, $M_{0(Abs)} = -8.93$ mag, corresponding to $m_{0(app)} = 32.34$ mag. The models illustrate that the observed distribution is consistent with an underlying turnover near this value, despite the lack of faint-end coverage in the data.}
  \label{fig:gclf}
\end{figure}


To ensure a robust and unbiased measurement of the GCLF in MACS0416, we restrict our analysis to sources that meet the 80\% completeness limit. This selection minimizes biases introduced by the rapid decline in detection efficiency at fainter magnitudes, where completeness corrections become increasingly uncertain. By limiting the sample in this way, we also reduce contamination from spurious detections and avoid over-reliance on model-dependent corrections, while still recovering the basic shape of the GCLF. Moreover, this approach follows standard practices in the literature, facilitating direct comparisons with previous studies \citep[\eg][]{harris2009catalog, villegas2010globular, jennings2014globular, alamo2017formation}. Here, the apparent magnitudes have been converted into absolute magnitudes, computed as $M_{200} = m_{F200W} - (m - M)_0 - K_{200} - A_{200}$, where $m_{F200W}$ is the observed apparent magnitude, $(m-M)_0 = 41.73$ is the distance modulus, $K_{200}$ is the $K$-correction appropriate for GCs at $z=0.397$ (Table~\ref{tab:kcorr}), and $A_{200}$ is the Galactic extinction correction ($A_{200} = 0.013$ mag).

Figure~\ref{fig:gclf} shows the observed $F200W$ GCLF (gray points with Poisson uncertainties), compared to theoretical log-normal models (dashed) and their completeness-corrected counterparts (solid). The intrinsic GCLF is modeled as a log-normal distribution:
\[
\phi(m) = \frac{1}{\sigma \sqrt{2\pi}} \exp\left[ -\frac{(m - m_0)^2}{2\sigma^2} \right],
\]
\n where $m_0 = -8.92$ mag is the predicted absolute turnover magnitude based on empirical measurements from similar clusters observed with JWST \citep[e.g.,][]{harris2024jwst}. Figure~\ref{fig:gclf} shows varying $\sigma$ values to explore its effect on the shape of the observed distribution. The observational incompleteness, $f(m)$, is modeled using the scaled hyperbolic tangent function shown in Fig.\ref{fig:stats_plot} and multiplied by log-normal GCLF $\phi(m)$, with best-fitting parameters $\beta = 2.26$ and $m_1 = 30.63$ mag.  Although the observed data do not reach sufficiently deep to sample the GCLF turnover, the completeness-corrected models allow us to extrapolate the expected behavior, and show that an expected turnover in the GCLF near $m_0 = -8.92$ mag is consistent with the observed bright-end slope and our detection limits.

To properly normalize the model GCLF, we apply a least-squares minimization across all magnitude bins rather than scaling to the histogram peak. The normalization factor $A$ is chosen to minimize the chi-squared statistic:
\[
\chi^2 = \sum_i \frac{[N_{\mathrm{obs},i} - A \cdot N_{\mathrm{model},i}]^2}{\sigma_i^2},
\]
where $N_{\mathrm{obs},i}$ is the observed number of GCs in bin $i$, $N_{\mathrm{model},i}$ is the model prediction integrated over each 0.2-mag bin, and $\sigma_i$ is the uncertainty in each bin. Only bins with completeness $\geq 80\%$ are included in the fit. This approach yields a best-fit normalization of $A = (8.9 \pm 0.2)\times10^3$, a turnover magnitude of $m_0 = -8.92 \pm 0.30$ mag (prior-dominated), and a dispersion of $\sigma = 1.63 \pm 0.06$. These results favor a moderately broad intrinsic distribution and are consistent with expectations for evolved, old globular cluster populations in massive galaxies at intermediate redshift.

This estimate is consistent with previous measurements of GCLF turnovers in similarly distant clusters, after accounting for differences in distance and filter bandpass \citep[\eg][]{Faisst2022, Lee2022, harris2023jwst, harris2024jwst}. Furthermore, this result aligns with studies of GC systems in clusters at $z \sim 0.4$--$0.5$, which report mild brightening of the observed GCLF turnover in redder or infrared filters, attributable to passive evolution and bandpass shifting \citep[\eg][]{jee2009cl0024, alamo2017formation}. Our findings support the interpretation that MACS0416 hosts a mature GC population whose LF closely resembles those of local systems when accounting for stellar population aging and filter effects.

While the observed sample captures the bright-end rise of the GCLF, it does not extend faint enough to directly sample the expected turnover at $m_0 \approx 31.04$ mag. Based on the intrinsic log-normal models shown in Figure~\ref{fig:gclf}, only $\sim$1–5\% of GCs are expected to be brighter than $m_{F200W}=30$. This implies that on the order of a few $\times10^{3}$ additional, fainter clusters would be required to fully sample the turnover and declining faint end of the luminosity function. These missing data points would span the magnitude range $30 \lesssim m \lesssim 32.5$ mag, where the current sample becomes increasingly incomplete. This estimate assumes a completeness level of at least 90\% across that range, which is critical for reliably constraining the full shape of the GCLF \citep[e.g.,][]{rejkuba2012globular, villegas2010globular, harris2014globular}. Although our current observations do not reach this depth, the completeness-corrected models provide a plausible extrapolation of the intrinsic distribution and support the expected turnover location \citep[\eg][]{harris2023jwst, harris2024jwst}.

\section{GC Mass Scales, UCDs, and Stellar Population Models}

Although the MACS0416 photometry does not reach the expected turnover magnitude of the GCLF, the measured distribution still provides valuable insight into the bright end of the cluster population. In this section, we extend the analysis beyond the GCLF to investigate the physical characteristics of these systems, including their stellar-mass scales, possible connection to ultra-compact dwarfs (UCDs), and consistency with single-stellar-population (SSP) models. Together, these analyses help to place the MACS0416 GC candidates within the broader context of GC and compact-stellar-system populations observed in nearby galaxies.

\subsection{GC Mass Scales and UCD Comparison} \label{sec:ucd_comparison}

In addition to the luminosity–function analysis, we constructed a $K$-corrected color–magnitude diagram (CMD) using $F200W_{ABS}$ versus $(F150W{-}F200W)_0$, chosen for consistency with the band employed in the GCLF study. 
To provide a stellar-mass scale for comparison with UCDs, we overplotted reference lines corresponding to $10^7$ and $10^8\,M_\odot$. 
These values were derived assuming an old single-stellar-population model with $(M/L)_{200}=1.5$ and a solar absolute magnitude of $M(F200W)_{\odot}=+5.3$\,AB-mag, consistent with standard population-synthesis models \citep[e.g.,][]{BruzualCharlot2003, Conroy2010} and empirically calibrated GC mass-to-light ratios from dynamical studies \citep{Kruijssen2008, Kruijssen2009}. 
The precise location of these mass lines varies by only $\pm0.3$\,mag across a plausible $(M/L)$ range of 1.0–2.0.  See also \citet{Harris2025} for similar model comparisons with five other GC systems at different redshifts.

As shown in Figure~\ref{fig:cmd_models}, the majority of GC candidates occupy the mass range $10^7\!<\!M/M_\odot\!<\!10^8$, consistent with the high-mass tail of the classical GC population \citep{harris2014globular,harris2023jwst}.  
Only a few luminous sources lie above $10^8\,M_\odot$, suggestive of possible UCD-like systems \citep[e.g.,][]{Hilker2009, Mieske2012, Norris2014}. 
While some contribution from UCD-like objects cannot be excluded, the bright-end distribution in MACS0416 is overwhelmingly dominated by classical GCs.

These mass estimates are luminosity-based and therefore depend on the assumed stellar-population age: older populations yield higher inferred masses for a given luminosity \citep[e.g.,][]{Hopkins2018}. 
At the redshift of MACS0416 ($z\!\approx\!0.397$), the GCs are expected to be several Gyr younger than present-day Galactic clusters, implying lower intrinsic $(M/L)$ ratios and hence higher apparent luminosities. 
This age effect could shift the observed distribution upward by $\sim$0.3–0.5\,dex in luminosity for a given true mass, reinforcing that most sources in Figure~\ref{fig:cmd_models} remain consistent with the classical GC regime even if they appear brighter than old-population models predict.

\subsection{Stellar Population Models} \label{sec:stellar_models}

To test whether the observed colors and magnitudes of the MACS0416 GC candidates are consistent with expectations for classical clusters at this redshift, we compared the data with single–stellar–population (SSP) models from the PARSEC v1.2S isochrone library \citep{Bressan2012, Marigo2017}. 
The models were generated for a grid of ages from $5$ to $9$\,Gyr (in 1\,Gyr steps) and for two representative metallicities, $[{\rm M/H}] = -2.0$ (metal–poor) and $+0.2$ (metal–rich). 
Model magnitudes were converted from Vega to AB using the JWST/NIRCam zeropoints, corrected for Galactic extinction, and $K$–corrected to the redshift of MACS0416 ($z = 0.397$).

\begin{figure}
  \centering
  \includegraphics[width=0.6\hsize]{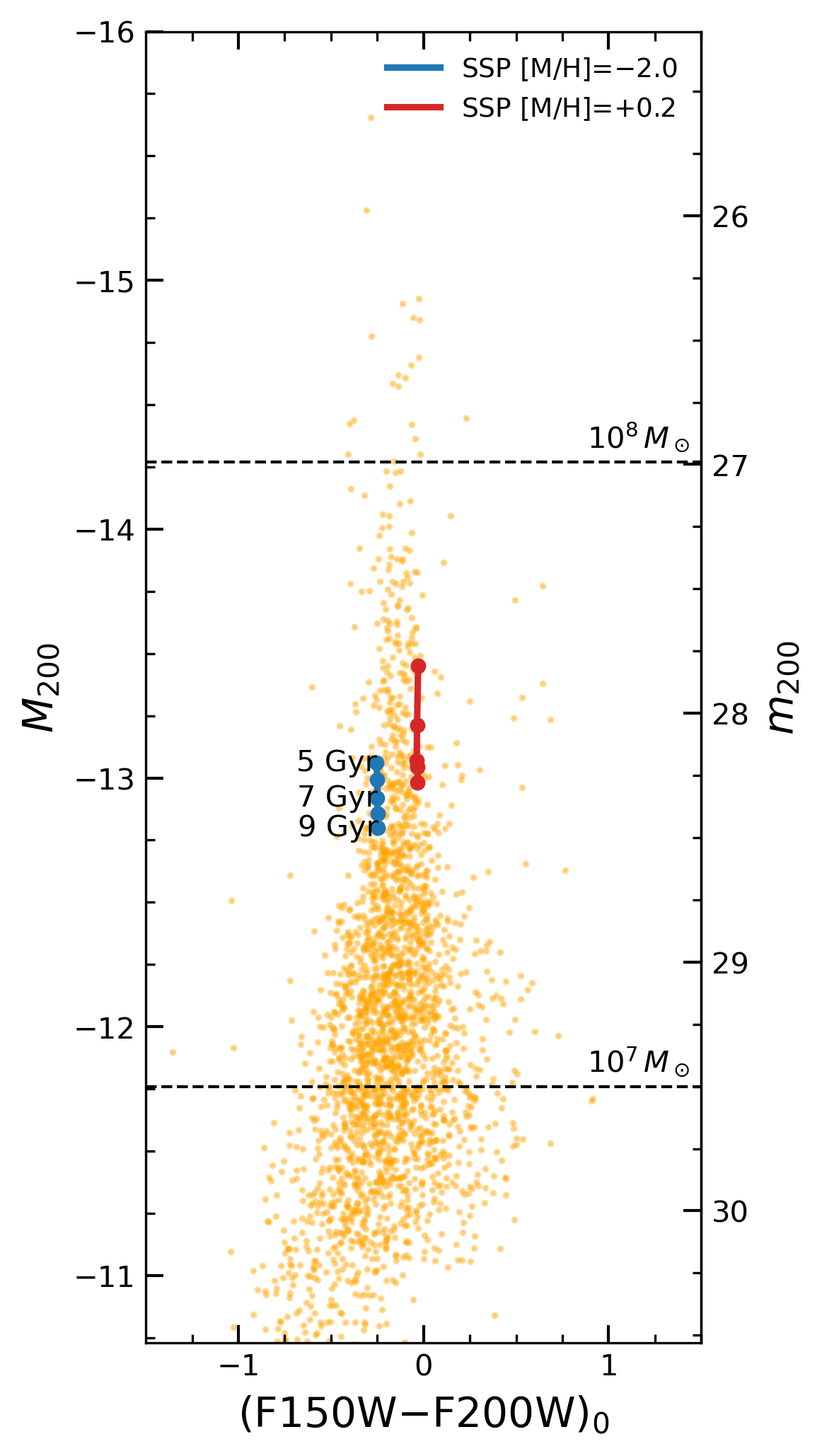}
  \caption{
  $K$-corrected CMD of MACS0416 GCs in $M_{F200W}$ versus $(F150W{-}F200W)_0$. 
Dashed lines mark stellar-mass reference levels of $10^7$ and $10^8\,M_\odot$, derived for $(M/L)_{F200W}=1.5$ and a solar $M(F200W)_{\odot}=+5.3$ AB-mag. 
PARSEC SSP tracks for $[{\rm M/H}]=-2.0$ (blue) and $+0.2$ (red) are shown with filled symbols denoting ages from 5–9 Gyr. 
Most candidates fall between the $10^7$–$10^8\,M_\odot$ lines and within the 5–9 Gyr loci, consistent with an intermediate-to-old GC population, while a few bright objects above these thresholds may correspond to UCD-like systems or modest age/metallicity variations.}
  
  \label{fig:cmd_models}
\end{figure}

Figure~\ref{fig:cmd_models} shows the resulting color–magnitude diagram with the SSP sequences overlaid. 
The blue and red lines represent the metal–poor and metal–rich models, respectively, with filled symbols marking ages from $5$ to $9$\,Gyr, increasing age from top to bottom. 
Across this range, the models predict the modest blueward color evolution and brightening expected for younger stellar populations. 
The majority of GC candidates lie between these sequences and are bracketed by the $5$–$9$\,Gyr loci, indicating a metallicity spread and age distribution consistent with intermediate–to–old GC systems observed at similar look–back times \citep[e.g.,][]{Harris2025}. 
This behavior parallels that seen in the MACS0417 cluster, where the GC population also follows PARSEC isochrones of comparable age and metallicity, reinforcing the interpretation that MACS0416 hosts a typical, evolved GC system rather than a predominantly young or star–forming population.

As noted above, small number of sources appear slightly above the $5$–$9$\ Gyr model sequences, which may reflect photometric scatter, modest age or metallicity variations, or the presence of a few objects more consistent with ultra-compact dwarfs or the remnant nuclei of tidally stripped galaxies. 
Overall, the observed distribution is well reproduced by standard SSP predictions across this age range, indicating that the bulk of the MACS0416 candidates are consistent with classical GCs at $z \simeq 0.4$, while a handful of luminous outliers may represent the transition toward the UCD regime discussed in Section~6.1.

\section{Discussion}

An additional point worth noting is the dynamical state of MACS0416 itself. Multiple studies of its lensing mass distribution and galaxy population indicate that MACS0416 is an ongoing merger, with an elongated mass morphology and two dominant subclusters \citep[e.g.,][]{grillo2015clash, Jauzac2016, diego2023jwst}. Such environments are expected to influence the globular cluster population by tidally stripping GCs from cluster galaxies, redistributing them into an intracluster component, and dynamically heating their spatial distribution \citep[e.g.,][]{Bekki2006}. In principle, this could enhance the number of intracluster GCs and broaden the overall spatial profile relative to more dynamically relaxed clusters. Our current data reach only the bright end of the GCLF, but the fitted turnover and dispersion are consistent, within uncertainties, with those of non-merging clusters at comparable redshifts \citep[e.g.,][]{harris2014globular, villegas2010globular}. This suggests that any merger-induced effects on the global LF are modest at the present sensitivity, though a more detailed radial analysis may reveal differences in the spatial distribution or intracluster fraction.

In future follow-up work, the MACS0416 GC population will be studied in greater detail. Quantifying how GC density declines with projected radius from the cluster center will help differentiate between GC subpopulations (e.g., intracluster vs.\ galaxy-bound GCs) and test for truncation or excesses due to dynamical stripping. Using modern simulations such as EMT-Pathfinder \citep{reina2024rescuer} to create mock JWST observations of GCs in the MACS0416 environment, incorporating realistic ages, metallicities, stellar evolution tracks, and observational effects, will provide an important supplement to this study and enable more direct comparisons between theory and observations.

\section{Summary and Conclusions}

In this study, we have used deep PEARLS mosaic images of MACS0416 taken with JWST/NIRCam to build a catalog of nearly 3,000 point-like sources surrounding the central galaxy cluster. Photometry in the NIRCam filters F090W, F115W, F150W, and F200W was performed using PSF fitting with \textsc{DAOPHOT}, with all four filters reaching comparable photometric depths. Empirical PSFs were characterized across the mosaics and verified to match the theoretical JWST/NIRCam profiles within 5–10\%, ensuring uniformity across the field. Recovery probability and internal measurement uncertainties were quantified through extensive artificial-star tests. A summary of our main findings is as follows:

\begin{enumerate}[label=\arabic*)]
    \item Photometry was performed with \textsc{DAOPHOT} in IRAF, using a stacked ultradeep image for source detection. Compact GCs in MACS0416 appear unresolved at JWST resolution. Final sources were required to be matched in all four filters and filtered using the \emph{sharp} parameter to remove non-stellar contaminants. A bright-end magnitude cut at $m_{F200W} < 26$ was adopted to separate potential UCDs and stripped nuclei from the GC sample.
    
    \item Artificial-star tests using the \textsc{addstar} function were used to evaluate completeness, revealing that the recovery probability remains above 80\% for sources brighter than $F200W \approx 30.36$~mag, with a 50\% completeness limit at 30.63~mag. This was modeled using a modified hyperbolic tangent function to quantify detection efficiency across magnitudes.
    
    \item CMDs of detected point sources show increasing scatter and declining completeness at fainter magnitudes, highlighting the importance of accounting for completeness when analyzing GC properties near the detection limit.
    
    \item Statistical analysis with the KMM algorithm sets limits on the possible presence of multiple GC subpopulations in MACS0416, particularly in the $(F115W{-}F200W)$ color. No clear visual evidence of distinct ``bimodal'' peaks is found, although a broad color spread is present.
    
    \item By restricting the sample to sources above the 80\% completeness limit, the luminosity-function (LF) analysis avoids biases from faint-end incompleteness, while still recovering the bright-end shape of the GCLF and enabling meaningful model comparisons.
    
    \item Although the data do not reach the expected turnover magnitude near $M_{0,\mathrm{abs}} = -8.93$~mag ($m_{0,\mathrm{app}} = 32.34$~mag), completeness-corrected log-normal models and comparisons with other clusters at similar redshift support the presence of a mature, evolved GC population in MACS0416.
    
    \item A $K$-corrected CMD incorporating stellar-mass reference lines and PARSEC SSP models shows that nearly all candidates are consistent with classical GCs below $10^7\,M_\odot$, with only a handful of brighter sources extending into the UCD regime. These luminous outliers are plausibly associated with ultra-compact dwarfs or the remnant nuclei of tidally stripped galaxies.
\end{enumerate}

Overall, MACS0416 provides a unique view of the globular-cluster population in a massive, intermediate-redshift cluster environment. Although the current JWST data probe only the bright end of the GCLF, they place strong constraints on the LF shape, color distribution, completeness, and contamination levels. The comparison with SSP models indicates that most MACS0416 GCs are intermediate-to-old systems consistent with a well-evolved population at $z \simeq 0.4$. Future JWST observations that reach fainter limits, coupled with multiwavelength optical–near-IR baselines, will be crucial for tracing the full GCLF turnover and disentangling GC subpopulations. Expanding similar analyzes across multiple lensing clusters will ultimately refine our understanding of GC formation, evolution, and the connection between massive clusters, UCDs, and stripped nuclei over cosmic time.

\begin{acknowledgments}
 This work is based on observations made with the NASA/ESA/CSA James Webb Space Telescope. All of the data presented in this article were obtained from the Mikulski Archive for Space Telescopes (MAST) at the Space Telescope Science Institute. The specific observations analyzed can be accessed via \textbackslash dataset \ [doi: 10.17909/qydg-zt87]{[https://doi.org/10.17909/qydg-zt87}](https://doi.org/10.17909/qydg-zt87). These observations are associated with JWST programs 1176 and 2738. R.A.W. acknowledges support from NASA JWST Interdisciplinary Scientist grants NAG5-12460, NNX14AN10G and 80NSSC18K0200 from GSFC. 
 This project was funded by the Agencia Estatal de Investigaci\'on, Unidad de Excelencia Mar\'ia de Maeztu, ref. MDM-2017-0765. J.M.B acknowledges support from the Beus Center for Cosmic Foundations at Arizona State University.
M.N. acknowledges INAF-Mainstreams 1.05.01.86.20. M.A.M. acknowledges the support of a National Research Council of Canada Plaskett Fellowship and the Australian Research Council Centre of Excellence for All Sky Astrophysics in 3 Dimensions (ASTRO 3D) through project number CE17010001. C.N.A.W. acknowledges funding from the JWST/NIRCam contract NASS-0215 to the University of Arizona.  W.E.H. acknowledges support from the Natural Sciences and Engineering Research Council of Canada through a Discovery Grant.
\end{acknowledgments}

{\it Facilities}: Hubble and JWST Mikulski Archive \url{https:// archive.stsci.edu}. Our specific GTO PEARLS observations were retrieved from MAST at STScI, and can be accessed via the following data sets: \dataset[10.3847/1538-3881/aca163].

{\it Software}: Astropy: \url{http://www.astropy.org} (Astropy Collaboration et al. 2013, 2018); \textsc{DAOPHOT} \citep{stetson1987DAOPHOT}, IRAF \citep{tody1986iraf, tody1993iraf}, Jupyter Notebook \citep{kluyver2016jupyter}, Matplotlib \citep{hunter2007matplotlib}, Numpy \citep{harris2020numpy}

\bibliography{bibliography}
\appendix 

\section{Contamination Estimation and Correction} \label{sec:appendix_contam}

Comparison of spatial source density before and after contamination correction. Each panel shows the distribution of detected point sources overlaid with grid cells used for contamination estimation. Grid cells are annotated with the remaining source count after applying the grid-based random removal procedure. By averaging the number of sources in 10 cells in the lower left corner, we find approximately 10 contaminating objects per cell.

\begin{figure}[htbp]
\centering
\includegraphics[width=1\textwidth,height=\textheight,keepaspectratio]{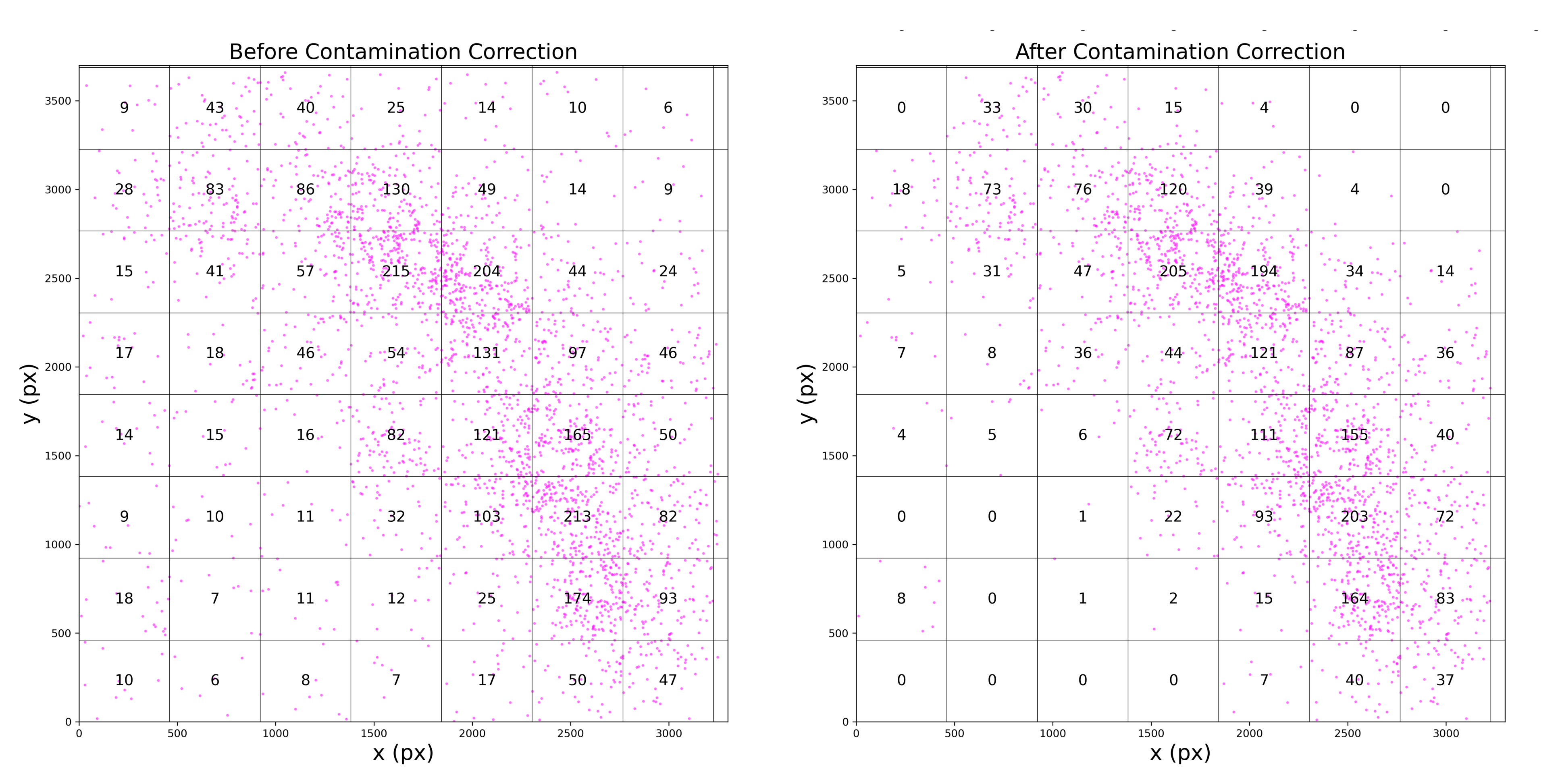}
\caption{\textit{Left:} Before contamination correction. \textit{Right:} After contamination correction. Note that the background estimate of $\sim$10 sources per cell is derived \emph{after} applying the same GC candidate selection criteria (multi-filter detection, sharpness/roundness limits, and PSF-based classification) to both the central cluster field and the outer pointings, ensuring a consistent comparison.}
\label{fig:contam_grid}
\end{figure}

\section{}

\end{document}